\documentstyle[12pt]{article}
\title{Bose-Einstein correlations in high energy multiple particle production
processes}
\author{Kacper Zalewski\thanks{Supported in part by the KBN grant
2P03B 086 14}\\
Jagellonian University\\ and\\ Institute of Nuclear Physics, Krak\'ow, Poland}

\begin{document}
\maketitle

\begin{abstract}
Correlations among identical bosons, which are familiar from statistical
physics, play an increasingly important role in high energy multiple particle
production processes. They provide information about the region, where the
particles are produced and, if Einstein's condensation can be reached, they can
lead to spectacular new phenomena.
\end{abstract}

\section{INTRODUCTION}

In this paper we will consider Bose-Einstein correlations in high-energy
particle production processes, i.e. the correlations among identical bosons in
the final state, which follow from Bose-Einstein statistics. When hundreds of
identical bosons are being produced in a single scattering act, as happens e.g.
in heavy ion collisions at high energy, such correlations can lead to
spectacular phenomena. They are also, most probably, the best way of getting
information about the space-time structure of the region, where the final state
particles are produced. Let us begin with a very simple example.

Consider the elastic scattering of two alpha particles with initial momenta
equal in magnitude, opposite and parallel to a horizontal axis, say the
$x$-axis. Suppose that the detectors register the final state particles if and
only if the scattering is at $90^\circ$ and one of the final particles goes up
and hits the upper detector ($U$), while the other goes down and hits the lower
detector ($L$). There are two possibilities. Either particle $1$, say the
particle coming from the left, hits detector $U$ and particle $2$ hits
detector $L$, or particle $1$ hits detector $L$ and particle $2$ hits detector
$U$. Let us denote the probability amplitudes for these two processes by $A$
and $B$ respectively. Since a rotation around the $x$-axis can convert these
two processes into each other, $|A| = |B|$. Since the alpha particles are
identical bosons and an exchange of the two final state particles converts the
two processes into each other, $A = B$. The detection probability is $|A + B|^2
= 4 |A|^2$. If the particles were distinguishable, the probability would be
$|A|^2 + |B|^2 = 2 |A|^2$. Thus, the fact that the particles are
indistinguishable increases the probability by a factor of two.

Let us make a few comments about this simple result.
\begin{itemize}

\item In the example the two amplitudes interfere constructively, because they
are coherent. This is sometimes called first order interference. We will see in
the following that the Bose-Einstein correlations of interest for us are due to
the incoherence of the production process, and are a manifestation of the so
called second order interference.

\item The statement that the scattering probability for identical particles is
twice the corresponding probability for distinguishable particles is not
possible to check experimentally, because non identical alpha particles are not
available. The best one can do is to compare the experimental result for the
identical alpha particles with the calculation for the non identical ones. In
the present example, where the calculation is simple and non controversial,
this is not much of a problem, but in multiple production processes a
calculation from first principles is not possible and the definition of the
distribution for distinguishable particles, which should be modified by the
Bose-Einstein correlations to yield the distribution which can be compared with
experiment, is a difficulty.

\item The final state can be represented by the density operator

\begin{equation}
\hat{\rho} = \frac{1}{2}|U_1 L_2 + U_2 L_1\rangle \langle U_1 L_2 + U_2 L_1|,
\end{equation}

where $U_i L_k$ is the state, where particle $i$ is registered by the detector
$U$ and particle $k$ by detector $L$. Expanding the left-had-side one obtains
four terms, if, however, the density operator is to be used only for
calculating averages of operators symmetric with respect to exchanges of the
identical particles, which is sufficient for all practical applications, one
can use the simpler form

\begin{equation}
\hat{\rho} = |U_1L_2\rangle\langle U_1L_2| + |U_1L_2\rangle \langle U_2L_1|.
\end{equation}

It is useful to rewrite this formula in the form

\begin{equation}
\hat{\rho} = \sum_P |U_1L_2\rangle \langle U_{P1}L_{P2}|,
\end{equation}

where the summation is over all the permutations P of the indices $1$ and $2$
and $Pi$ denotes the index obtained from index $i$ under permutation $P$. For
our simple example this formula is ridiculously complicated, for more
difficult cases, however, its analogues are very convenient.

\end{itemize}

\section{HBT contribution}

An interesting application of the Bose-Einstein interference to find the sizes
of the emitting objects was discovered in the fifties by two astronomers, R.
Hanbury Brown and R.Q. Twiss \cite{HBT}. By studying the second order Bose-
Einstein interference of photons, they were able to measure the radii of some
stars. The idea may seem obvious today, but it was not so at the time it was
put forward. In the seventies Hanbury Brown wrote \cite{HBR} (quoted after
\cite{SIL})
\begin{quote}
\em Now to a surprising number of people this idea seemed not only heretical
but patently absurd and they told us so in person, by letter, in publications,
and by actually doing experiments which claimed to show that we were wrong. At
the most basic level they asked how, if photons are emitted at random in a
thermal source can they appear in pairs at the two detectors. At a more
sophisticated level the enraged physicist would brandish some sacred text,
usually by Heitler, and point out that \ldots our analysis was invalidated by
the uncertainty relation \ldots
\end{quote}

Since the distances to many stars are known, their radii could be determined,
if the opening angles between the light rays coming from the stars could be
measured. These angles, however, are in most cases too small for a direct
measurement. Hanbury Brown and Twiss suggested the following procedure.
Consider two light rays coming from two points on the surface of the star --
ray $a$ from point $a$ and ray $b$ from point $b$. The problem is to measure
the angle $\Theta$ between the two rays. Each of the rays falls on two
photodetectors denoted $1$ and $2$. The distance between the photodetectors is
$d$. Elementary trigonometry yields to first order in $\Theta$ the relation

\begin{equation}
\Theta = \frac{\Delta_a - \Delta_b}{d \sin\alpha},
\end{equation}
where $\Delta_i$ is the difference of distances between point $i$ on the star
and the two photodetectors, while $\alpha$ is the angle between the line
connecting the two photodetectors and the direction of the two rays. Thus the
problem of measuring the opening angle $\Theta$ reduces to the problem of
measuring $\Delta_a - \Delta_b$. Of course in practice, in order to find the
radius of the star a suitable averaging over the possible emission points is
necessary, this, however, is rather simple and we shall not discuss it any
further.

The current generated in the photodetector is proportional to the intensity of
the incident light. Thus for counter $1$ it is

\begin{equation}
i_{1u} = K_1\left[E_a\sin(\omega_a t+\phi_a) + E_b\sin(\omega_b t+\phi_b)
\right]^2,
\end{equation}
where $\phi_i,\;E_i,\;\omega_i$ denote respectively the phase at the star
surface, the amplitude and the frequency for ray $i$, and $K_1$ is a
proportionality coefficient dependent on the working of the photodetector $1$.
For simplicity, the polarization effects have been ignored and the time
necessary to reach photodetector $1$ has been put equal $t$ for both point $a$
and point $b$. In the apparatus the current $i_{1u}$ is further filtered so that
only frequencies from 1 Hz to 100 Hz survive. Thus finally, the current from
the first photodetector is

\begin{equation}
i_1 = K_1 E_a E_b \cos[(\omega_a-\omega_b)t + (\phi_a-\phi_b)].
\end{equation}
This current is zero on the average and does not look particularly interesting.
The analysis for the second photodetector is similar except that the time
necessary to reach the detector for the ray from point $i$ on the star is
increased by $\Delta_i/c$ . One obtains

\begin{equation}
i_2 = K_2 E_a E_b \cos\left[(\omega_a - \omega_b)t + \frac{\omega}{c} (\Delta_a
- \Delta_b) + (\phi_a - \phi_b) \right],
\end{equation}
where $\omega \approx \omega_a \approx \omega_b$. This is again a rather
uninteresting current, but the average of the product of the filtered
currents from the two photodetectors

\begin{equation}
\langle i_1 i_2 \rangle = K_1 K_2 E_a E_b \cos\left[ \frac{\omega}{c} (\Delta_b
- \Delta_a)\right],
\end{equation}
which is measurable, yields $\Delta_a - \Delta_b$ and consequently the
necessary opening angle $\Theta$.

Note that the result is obtained in spite of the fact that the presence of the
random phases $\phi_i$ means that light from $a$ is incoherent with respect to
light from $b$. Because of these phases the product of two amplitudes, one for
the ray $a$ and one for the ray $b$ averages to zero, The product of four
amplitudes, two from $a$ and two from $b$, however, can survive. For this
reason one calls this effect second order interference or intensity
interferometry.

\section{The GGLP contribution}

The first application of intensity interferometry in particle physics was made
by the Goldhabers Lee and Pais \cite{GGL}. Their problem was somewhat different
from that that in the HBT case. The interfering particles were like sign pion
produced at two points within the interaction region of a hadron - hadron
collision. The interference of interest was not between the measurements at two
points in space, but between momentum measurements. Assuming that at the
production space-time point $x_k$ the pion wave function has phase $\phi_k$ and
that the momentum of the pion $p_k$ is well-defined, one expects at the
registration point $x$ an amplitude proportional to $ \exp[ ip_k (x_k-x) + i
\phi_k]$. The probability of finding the two pions produced at points $x_1$ and
$x_2$ with momenta $p_1$ and $p_2$, after proper symmetrization of the wave
function, is proportional to

\begin{equation}
\frac{1}{2}\left| e^{i(p_1x_1 + p_2x_2)} + e^{i(p_1x_2 + p_2x_1)} \right|^2 = 1
+ \cos[(p_1-p_2)(x_1-x_2)].
\end{equation}
We assume now that the production process is incoherent, so that the averaging
over the times and positions $x_1,\;x_2$ should be made at the level of
probabilities and not of amplitudes. Then the distribution of the difference in
momenta should be approximately given by the formula

\begin{equation}
C(p_1 - p_2) = 1 + \langle \cos[(p_1-p_2)(x_1-x_2)] \rangle,
\end{equation}
where the averaging is over $x_1$ and $x_2$. Qualitatively, the result does
not depend much on the actual prescription being used for the averaging. For
$p_1 \approx p_2$ the argument of the cosine is close to zero and consequently
$C \approx 2$. For large momentum differences, the argument of the cosine is
a rapidly oscillating function of $x_1-x_2$, which is strongly suppressed by
the averaging process, and $C \approx 1$. If the weight function used for the
averaging contains just one parameter with the dimension of length, let us
denote it R, the width of region in $p_1-p_2$, where $C$ is significantly
bigger than one, must be of the order of $R^{-1}$. There are many specific
recipes how to perform the averaging. The results obtained for the
correlations of momenta and for the sizes and shapes of the interaction regions
are reasonable. For reviews see \cite{BGJ} and \cite{HAY}. In spite of this
success many difficulties remain.

\begin{itemize}

\item Since the energy of a pion is determined by its momentum, one has data
only on the three-dimensional distribution of the differences of spacial
momenta. This is not enough to derive the four-dimensional distribution of the
sources in space-time. Therefore, the results are strongly model dependent.

\item The averaging over the square of the wave function corresponds to the
assumption that the density matrix of the final pions in coordinate
representation is diagonal. This in turn implies that the momentum distribution
should be flat, which contradicts experiment. A closely related question is,
how the pion can be initially localized at the production point and then
represented by a plane wave corresponding to well-defined momentum.

\item Information about the production region is, in practice, obtained only
from pairs of pions with similar momenta. Consequently, what is being measured
is not the whole interaction region, but the region, where the pions with
similar momenta are produced. This, incidentally, explains the fact that the
interaction regions usually come out roughly spherical, while one believes that
the full interaction region is more string like.

\item There are many corrections, which probably should be applied, but it is
controversial how. Here belong the corrections for coulomb repulsion between
the charged like sign pions, the corrections for the final state interactions
due to strong coupling, the corrections due to resonance production, the
corrections due to partial coherence of the source etc.

\end{itemize}

\section{Density matrix approach}

In order to obtain a more general formulation for the GGLP problem it is
convenient to use the formalism of density matrices. This has been described
by a number of people, here we are the closest to the formulation used by
Bialas and Krzywicki \cite{BIK}. We introduce an auxiliary, unphysical process,
where all the particles produced are distinguishable. We assume that for this
process simple, intuitive ideas work. Then we correct for the Bose-Einstein
correlations in order to obtain results comparable with experiment. This
approach has its defects as discussed in the Introduction (see also
\cite{BGJ}, \cite{HAY}), but for lack of a better idea it is widely used. As
our starting point for the distinguishable particles we use an independent
production model (cf \cite{BZ1}, \cite{BZ2}, \cite{BZ3} and references
contained there).
In this model the multiplicity distribution for the particles is poissonian

\begin{equation}
P^{(0)}_N = \frac{\nu^N}{N!} e^{-\nu}
\end{equation}
and for each multiplicity the density matrix is a product of single particle
density matrices

\begin{equation}
\rho^{(0)}_N(q,q') = \prod_{i=1}^N \rho_1^{(0)}(q_i,q'_i).
\end{equation}
The momentum distribution is given, as usual, by the diagonal elements of the
density matrix in the momentum representation

\begin{equation}
\Omega_{0N}(q) = \rho_N^{(0)}(q,q).
\end{equation}
It is convenient to normalize it to unity

\begin{equation}
\sigma^{(0)}_N = \int\!\!dq\, \Omega_{0N}(q) = 1.
\end{equation}

For identical particles the density matrix should be symmetrized as explained
in the Introduction

\begin{equation}
\rho_N(q,q') = \sum_P \rho_N^{(0)}(q,q'_P).
\end{equation}
The corresponding momentum distribution for a given multiplicity is

\begin{equation}
\Omega_N = \rho_N(q,q).
\end{equation}
This, however, is no more normalized to unity, because

\begin{equation}
\sigma_N = \int dq\, \Omega_N(q) = 1 + \ldots
\end{equation}
The first term in the last expression corresponds to the identity permutation,
but there are $(N-1)!$ further terms. This yields the multiplicity distribution

\begin{equation}
P_N = {\cal N} P_N^{(0)} \sigma_N,
\end{equation}
where ${\cal N}$ is an $N$-independent normalizing factor, which ensures that
$\sum P_N = 1$.

\section{Simple case: pure final state}

In order to present simply the qualitative features of the result, let us
consider first the case, when for each multiplicity the final state is pure.
This is a grossly oversimplified model, but we will find that it contains some
features of the much more realistic approach presented in the following
section. For the pure state model

\begin{equation}
\hat{\rho}_N^{(0)} = |\psi_{N}^{(0)}\rangle \langle\psi_{N}^{(0)}|.
\end{equation}
We assume that each of the state vectors $|\psi_{N}^{(0)}\rangle$ is symmetric
with respect to exchanges of particles. Thus the effect of the summation over
the permutations $P$ is simply to multiply the operator $\hat{\rho}_{N}^{(0)}$
by $N!$. As a result the probability of producing exactly $N$ particles is also
multiplied by the factor $N!$. The Poisson distribution goes over into a
geometrical distribution and after evaluating the normalization factor we get

\begin{equation}
P_N = (1-\nu) \nu^N.
\end{equation}
This formula makes sense only if $\nu <1$, because otherwise the sum of the
probabilities $P_N$ diverges. For the average number of particles one finds

\begin{equation}
\overline{N} = \frac{\nu}{1 - \nu}
\end{equation}
with a singularity at $\nu = 1$. From the model presented in the following
section it will be seen that this singularity corresponds to Einstein's
condensation.

In order to avoid the repeated summation of series it is convenient to
introduce the generating functions. The generating function for the
multiplicity distribution is

\begin{equation}
\Phi(z) = \sum_{N=0}^{\infty} P_N z^N = \frac{1- \nu}{1 - z\nu}.
\end{equation}
The logarithmic derivative of this function with respect to $z$ at $z=1$ yields
the average multiplicity. The second derivative of the logarithm with respect to
$z$ at $z=1$ is the dispersion and in general the $p$-th cumulant of the
multiplicity distribution is given by

\begin{equation}
K_p = (p-1)! \left( \frac{\partial^p Log\Phi}{\partial z^p} \right)_{z=1} =
(p-1)!
\left( \frac{\nu}{1 - \nu} \right)^p.
\end{equation}

Inclusive and exclusive momentum distributions, as well as all the correlation
functions in momentum space, can be calculated by functional differentiation
from the generating functional

\begin{equation}
\Phi[u] = \sum_{N=0}^\infty {\cal N} P^{(0)}_N \int dq\,\Omega_N(q)
\prod_{i=1}^N u(q_i) = \frac{1-\nu}{1 - \nu \int dq_i \Omega_{01}(q_i)u(q_i)}.
\end{equation}
For instance, the single particle distribution is

\begin{equation}
\left( \frac{\delta \Phi[u]}{\delta u} \right)_{u=1} = \frac{\nu}{1-\nu}
\Omega_0(q).
\end{equation}
Thus symmetrization (Bose-Einstein statistics) introduces in this simple model
only a change of normalization.

\section{Independent production}

Let us consider now the full independent production model. In order to find the
modification of the multiplicity distribution due to Bose-Einsten statistics it
is necessary to calculate the correction factors

\begin{equation}
\sigma_N = \sum_P \int dq\, \prod_{i=1}^N \rho_1^{(0)}(q_i,q_{Pi}).
\end{equation}
Since each permutation can be decomposed into cycles, this integrals can be
expressed in terms of the cycle integrals

\begin{equation}
C_{k>1} = \int d^{3k}q\, \rho_1^{(0)}(q_1,q_2) \rho_1^{(0)}(q_2,q_3)\ldots
\rho_1^{(0)}(q_k,q_1).
\end{equation}
It is convenient to add the definition

\begin{equation}
C_1 = 1.
\end{equation}
Similarly the integrals necessary to calculate the generating functional for
the momentum distributions can be expressed in terms of the cycle integrals

\begin{equation}
C_k[u] = \int d^{3k}\,qu(q_1) \rho_1^{(0)}(q_1,q_2)u(q_2)
\rho_1^{(0)}(q_2,q_3)\ldots u(q_k)\rho_1^{(0)}(q_k,q_1)
\end{equation}

After some combinatorics, very similar to that used when deriving the linked
clusters expansion familiar from quantum field theory and many body theory, one
finds the generating functional

\begin{equation}
\Phi[u] = \exp\left[ \sum_{k=1}^\infty \nu^k \frac{C_k[u] - C_k[1]}{k} \right].
\end{equation}
Substituting in this functional $z$ for $u$ one obtains the generating function
for the multiplicity distribution. Without exhibiting the actual calculations we
will now present some general results, obtained for this model.

\begin{itemize}
\item The single particle momentum distribution and all the momentum
correlation functions can be expressed in terms of one function depending on
two single particle momenta

\begin{equation}
L(q_1,q'_1) = \sum_{k=1}^\infty \nu_k \int d^3q_2\ldots d^3q_k\,
\rho_1^{(0)}(q_1, q_2)
\rho_1^{(0)}(q_2, q_3) \ldots \rho_1^{(0)}(q_k, q'_1).
\end{equation}
For instance the momentum distribution is

\begin{equation}
\Omega(q) = L(q,q).
\end{equation}
The two particle cumulant is

\begin{equation}
K_2(q_1,q_2) = L(q_1,q_2)L(q_2,q_1).
\end{equation}
In general the $p$-th correlation function is

\begin{eqnarray}
K_p(q_1, q_2,\ldots,q_p) = L(q_1,q_2)L(q_2,q_3)\ldots L(q_p,q_1) + \nonumber \\
(\mbox{permutations of the indices  } 2,\ldots,p).
\end{eqnarray}
\item For typical density matrices the average square of the difference of
momenta between two particles $\langle q^2 \rangle$ decreases due to
symmetrization.
\item For typical density matrices the average difference between the
production points of pairs of particles decreases due to symmetrization.
\item For typical density matrices the size of the interaction region as
evaluated from the width of the two-particle momentum correlation function
decreases due to symmetrization.
\end{itemize}
We will discuss these predictions in a further section, where we will rederive
them in a more intuitive way. The references to "typical density matrices" mean
that the statement is true for most density matrices, but not for all. We have
not been able to find a condition defining the relevant class of density
matrices.

Probably the most interesting implication is the possibility of Einstein's
condensation, but this will be discussed in the following section.

\section{Einstein's condensation}

Using matrix notation one can rewrite the definition of the function $L$ given
in the previous section in the form

\begin{equation}
L(q,q') = \sum_{k=1}^\infty \nu^k \langle
q|\left(\hat{\rho}_1^{(0)}\right)^k|q'\rangle.
\end{equation}
Expanding the single particle density operator in terms of its eigenvectors and
eigenvalues, we find

\begin{equation}
\hat{\rho}_1^{(0)} = \sum_n |n\rangle \lambda_n \langle n|
\end{equation}
and for its $k$-th power

\begin{equation}
\left(\hat{\rho}^{(0)}_1\right)^k = \sum_n |n\rangle \lambda_n^k \langle n|.
\end{equation}
Thus in the momentum representation

\begin{equation}
L(q,q') = \sum_n \langle q|n\rangle \langle n| q' \rangle \sum_{k=1}^\infty
\nu^k \lambda_n^k.
\end{equation}
This expression makes sense only if for all $n$ there is $\lambda_n \nu <1$.
Denoting the largest eigenvalue of the density operator by $\lambda _0$, we
expect problems when $\nu\lambda_0 \rightarrow 1$.

Performing the summations of the geometric series, we can rewrite the
expression for $L(q,q')$ in the form

\begin{equation}
L(q,q') = \sum_n \frac{\psi_n(q) \psi_n^*(q') \nu \lambda_n}{1 - \nu\lambda_n}.
\end{equation}
For $\nu \lambda_0 \rightarrow 1$ it is convenient to use the equivalent
formula

\begin{equation}
L(q,q') = \frac{\psi_0(q)\psi_0^*(q')}{1-\nu\lambda_0} + \tilde{L}(q,q'),
\end{equation}
where $\tilde{L}$ remains bounded in the limit. Putting $q=q'$ and integrating
over $q$ we get the corresponding formula for the average multiplicity

\begin{equation}
\overline{N} = \frac{1}{1-\nu\lambda_n} + \mbox{bounded term}.
\end{equation}
From these formulae it is clear that when $\nu\lambda_0$ tends to one,
Einstein's condensation occurs. Increasing $\nu$ corresponds to the increasing
of the number of particles in the system. In all the states with indices $n
\neq 0$ there is place only for a limited number of particles, while all the
surplus, which can be arbitrarily large, gets located in the state $|0\rangle$.
In this sense, when the number of particles becomes very large, we recover the
model with the pure state discussed previously. A very interesting question is,
whether experimentally it is possible to create condition, where the Einstein
condensate would dominate.

Let us conclude this section with two remarks. For the Gaussian single
particle density matrix

\begin{equation}
\rho_1^{(0)}(q,q') = \frac{1}{\sqrt{2\pi\Delta^2}}
\exp\left[-\frac{q_+^2}{2\Delta^2}
- \frac{R^2}{2}q^2_-\right],
\end{equation}
where $q_+ = (q+q')/2$ and $q_- = q - q'$, the eigenvalues and eigenfunction
are known \cite{BZ1}. Thus all the calculations can be easily performed. In
fact they have been performed by various methods \cite{PR1}, \cite{PR2},
\cite{PR3}, \cite{CSZ}.

The theory can be reformulated in the second quantization formalism. Then the
function $L(q,q')$ appears as the Green function $\langle \hat{a}^{\dag}_q
\hat{a}_{q'} \rangle$ and the possibility of expressing all the correlation
functions in terms of $L(q,q')$ is the Wick theorem with $L$ as the only non
zero contraction.

\section{Statistical physics interpretation}

Many results from the previous sections can be simply reinterpreted and
rederived using standard statistical physics. Consider the single particle
unsymmetrized density operator

\begin{equation}
\hat{\rho}_1^0 = \sum_n |n\rangle \lambda_n \langle n|
\end{equation}
with the condition $\sum_n \lambda_n = 1$. This can be reinterpreted as the
density operator corresponding to the canonical ensemble, if we put

\begin{equation}
\lambda_n = \frac{1}{Z} e^{-\beta \varepsilon_n},
\end{equation}
where, as usual, $\varepsilon_n$ is the energy of state $|n\rangle$, $\beta$ is
the inverse temperature in energy units and $Z = \sum_n \exp[-\beta
\varepsilon_n]$ is the canonical partition function. The corresponding
(single particle) Hamiltonian is

\begin{equation}
\hat{H} = \sum_n|n\rangle \varepsilon_n \langle n|.
\end{equation}
This Hamiltonian, when written in the coordinate representation, may look quite
unusual, but some cases are simple. For instance, the Gaussian density matrix
corresponds to the Hamiltonian of a harmonic oscillator.

For indistinguishable particles a single particle is not a convenient subsystem
and, as suggested by Pauli long ago, it is better to choose as subsystem the
open system consisting of all the particles in state $|n\rangle$. The state of
this subsystem is defined by the number of particles $(N)$ in it. The
probability of state $N$ of the subsystem is

\begin{equation}
P_n(N) = \frac{1}{{\cal Z}_n} \overline{\nu}^N e^{-\beta N \varepsilon_n}.
\end{equation}
The grand partition function

\begin{equation}
{\cal Z}_N = \frac{1}{1 - \overline{\nu}e^{-\beta \varepsilon}}
\end{equation}
is chosen so that $\sum_{N=0}^\infty P_n(N) = 1$, the parameter
$\overline{\nu}$ is known in statistical physics as the fugacity and is
connected to the chemical potential $\mu$ by the formula

\begin{equation}
\overline\nu = e^{\beta\mu}.
\end{equation}
In order to reproduce the formulae from the previous sections, one puts
$Z\overline{\nu} = \nu$. The grand partition function can be used to find the
moments of the multiplicity distribution very much like the multiplicity
generating function. For instance, for the average occupation of state $n$ we
find

\begin{equation}
\langle N_n \rangle = -\frac{1}{\beta} \frac{\partial Log {\cal Z}_n}{\partial
\mu} = \frac{1}{e^{\beta(\varepsilon_n - \mu)} - 1} = \frac{\nu\lambda_n}{1 -
\nu\lambda_n}.
\end{equation}
The probability of no particles in state $n$ is

\begin{equation}
P_n(0) = \frac{1}{{\cal Z}_n} = 1 - \overline{\nu}e^{-\beta\varepsilon_n} =
\frac{1}{\langle N_n \rangle + 1}.
\end{equation}
The probability of no particle in the whole system is

\begin{equation}
P(0) = \prod_n \frac{1}{\langle N_n \rangle + 1}.
\end{equation}
Let us consider two limiting cases. When all the occupation numbers are very
small, the product equals approximately $\exp[-\langle N \rangle]$ and for
large multiplicities it is very small. Very large fluctuations of the
multiplicity are very unlikely to occur. When most particles are in the state
$n=0$, the product is approximately $(\langle N \rangle +1)^{-1}$, which is
much bigger than in the previous case. Thus, when there is much Einstein
condensate, large multiplcity fluctuations become much more probable. Cosmic
ray physicists have been reporting \cite{LFH} observations of centauro and
anticentauro events. This are high multiplcity events, where respectively
either the neutral pions or the charged pions are missing. One could speculate
that this phenomena are related to Einstein's condensation.

Statistical physics gives also a simple interpretation for the function
$L(q,q')$. One finds

\begin{equation}
L(q,q') = \sum_n \psi_n(q)\psi_n^*(q') \frac{1}{e^{\beta(\varepsilon_n - \mu)}
-
1}.
\end{equation}
This is the canonical density matrix with the Maxwell-Boltzmann weights
replaced by the Bose-Einstein weights. The fact that the Bose-Einstein weights
fall with increasing energy $\varepsilon_n$ faster than the Maxwell-Botzmann
weights explains qualitatively most of the observations reported previously.
For most Hamiltonians the wave function spreads in ordinary space and in
momentum space, when energy is increased. Since the Bose-Einstein weights
enhance the low energies, they reduce the average momenta and radii. Also the
reduction of the effective radius of the interaction region, as determined from
the width of the correlation function in momentum space, can be easily
understood. If in the previous formula all the terms had equal weights,
we would obtain $L(q,q') = \delta^3(q-q')$. The stronger the
cut on the sum, the broader the peak in $q-q'$ becomes. Since the Bose-Einstein
weights are more peaked at low energies than the Maxwell-Boltzmann ones, they
correspond to a broader peak in the correlation function. Since the width of
this peak is inversely proportional to the radius of the production region,
symmetrization reduces the radius of this region. All these qualitative
arguments are usually true. It is, however, easy to show examples of
hamiltonians, where e.g. with increasing energy the wave function shrinks
either in ordinary space, or in momentum space. Additional assumptions
necessary to convert these qualitative arguments into rigorous theorems are,
therefore, necessary, but not yet known.

\end{document}